\documentclass[10pt,twocolumn,letterpaper]{article}
\usepackage[pagenumbers]{cvpr} % To force page numbers, e.g. for an arXiv version

\usepackage{nicefrac}
\usepackage{pifont}
\usepackage{upgreek}
\definecolor{MyDarkBlue}{rgb}{0,0.08,1}
\definecolor{MyLightBlue}{RGB}{55, 140, 231}
\definecolor{MyDarkGreen}{rgb}{0.02,0.6,0.02}
\definecolor{MyDarkRed}{rgb}{0.8,0.02,0.02}
\definecolor{MyDarkOrange}{rgb}{0.40,0.2,0.02}
\definecolor{MyLightOrange}{rgb}{0.90,0.5,0.02}
\definecolor{MyPurple}{RGB}{111,0,255}
\definecolor{MyLightPurple}{RGB}{126, 96, 191}
\definecolor{MyRed}{rgb}{1.0,0.0,0.0}
\definecolor{MyGold}{rgb}{0.75,0.6,0.12}
\definecolor{MyDarkgray}{rgb}{0.66, 0.66, 0.66}
\definecolor{MyLightGray}{rgb}{0.8, 0.8, 0.8}
\definecolor{MyWineRed}{rgb}{0.694,0.071, 0.149}
\definecolor{nicegreen}{rgb}{0.1, 0.6, 0.2}

\newcommand{\secref}[1]{Sec.~\ref{#1}}
\newcommand{\figref}[1]{Fig.~\ref{#1}}
\newcommand{\myeqref}[1]{Eq.~\ref{#1}}

\DeclareMathOperator{\LBS}{\mathrm{LBS}}

\newcommand{\B}[0]{\mathcal{B}}
\newcommand{\G}[0]{\mathcal{G}}

\newcommand{\R}[0]{\mathbb{R}}
\newcommand{\MLPx}[0]{\mathcal{MLP}_{\chi}}
\newcommand{\MLPt}[0]{\mathcal{MLP}_{\theta}}
\newcommand{\hyper}[0]{\mathit{hyper}}
\newcommand{\Hyper}[0]{\xi_\hyper}
\newcommand{\psih}[0]{\psi_\hyper}

\newcommand\blfootnote[1]{%
  \begingroup
  \renewcommand\thefootnote{}\footnote{#1}%
  \addtocounter{footnote}{-1}%
  \endgroup
}

\definecolor{cvprblue}{rgb}{0.21,0.49,0.74}
\usepackage[pagebackref,breaklinks,colorlinks,allcolors=cvprblue]{hyperref}
\usepackage[table]{xcolor}
\title{PhySkin: Physics-based Bone-driven Neural Garment Simulation}
\author{%
Astitva Srivastava$^{*12}$ \hspace{0.3em}
Hsiao-yu Chen$^{2}$  \hspace{0.3em}
Ryan Goldade$^{2}$  \hspace{0.3em}
Philipp Herholz$^{2}$  \hspace{0.3em}
Zhongshi Jiang$^{2}$  \\
Gene Wei-Chin Lin$^{2}$ \hspace{0.3em}
Lingchen Yang$^{2}$ \hspace{0.3em}
Nikolaos Sarafianos$^{2}$ \hspace{0.3em}
Tuur Stuyck$^{2}$ \hspace{0.3em}
Egor Larionov$^{2}$ \hspace{0.3em} \\
{\normalsize $^1$IIIT Hyderabad, India}  \quad
{\normalsize $^2$Meta Reality Labs}
}

\begin{document}
\twocolumn[{%
\renewcommand\twocolumn[1][]{#1}%
\maketitle
\vspace{-9mm}
\centering
\includegraphics[width=0.9\linewidth]{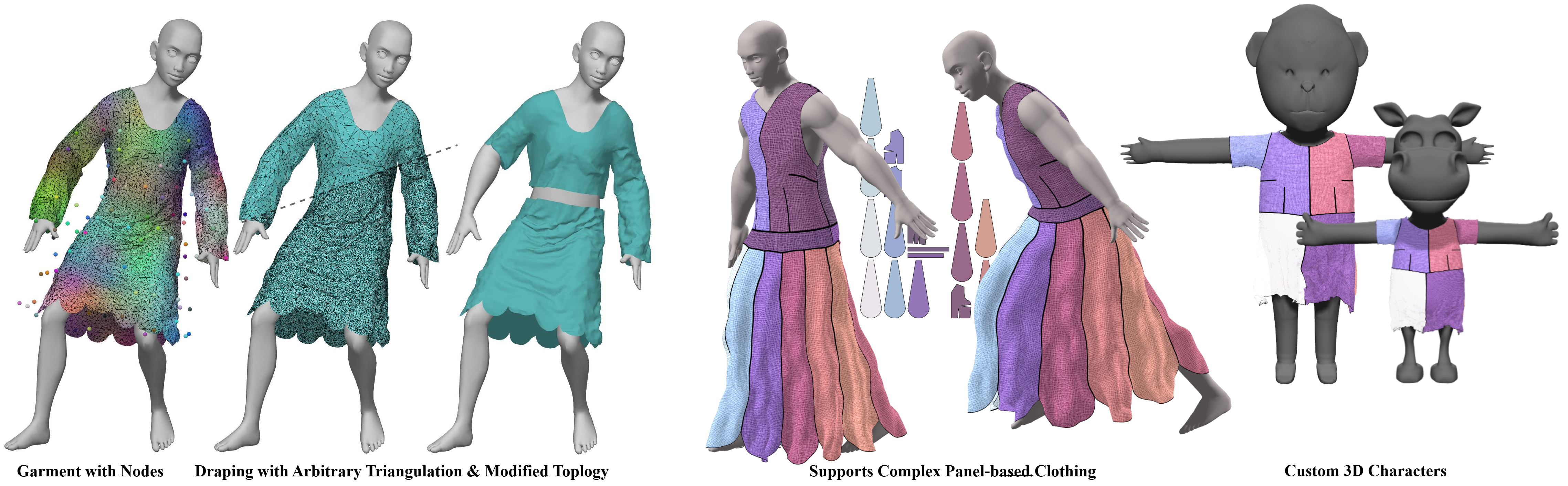}
\captionof{figure}{We propose a garment draping model that runs in microseconds using single-threaded CPU compute, which handles garments across a range of poses and body shapes. 
Our deformation approach performs physics-based quasi-static simulation in reduced subspaces defined by node-based transformations, to which the garment is skinned. 
The model is trained in a self-supervised manner and supports topology and resolution-independent inference.}
\label{fig:1_teaser}
\vspace{0.17cm}
}]

\begin{abstract}
Recent advances in digital avatar technology have enabled the generation of compelling virtual characters, but deploying these avatars on compute-constrained devices poses significant challenges for achieving realistic garment deformations. 
While physics-based simulations yield accurate results, they are computationally prohibitive for real-time applications. Conversely, linear blend skinning offers efficiency but fails to capture the complex dynamics of loose-fitting garments, resulting in unrealistic motion and visual artifacts. Neural methods have shown promise, yet they struggle to animate loose clothing plausibly under strict performance constraints.
In this work, we present a novel approach for fast and physically plausible garment draping tailored for resource-constrained environments. 
Our method leverages a reduced-space quasi-static neural simulation, mapping the garment’s full degrees of freedom to a set of bone handles that drive deformation. 
A neural deformation model is trained in a fully self-supervised manner, eliminating the need for costly simulation data. At runtime, a lightweight neural network modulates the handle deformations based on body shape and pose, enabling realistic garment behavior that respects physical properties such as gravity, fabric stretching, bending, and collision avoidance.
Experimental results demonstrate that our method achieves physically plausible garment drapes while generalizing across diverse poses and body shapes, supporting zero-shot evaluation and mesh topology independence.
Our method's runtime significantly outperforms past works, as it runs in microseconds per frame using single-threaded CPU inference, offering a practical solution for real-time avatar animation on low-compute devices. 
\blfootnote{*This work was conducted during an internship at Meta Reality Labs}
\end{abstract}
% Our approach generalizes across diverse poses and body shapes, supporting zero-shot evaluation and mesh topology independence. Experimental results demonstrate that our method achieves physically plausible garment drapes with significantly improved runtime performance compared to past techniques.     
\section{Introduction}\label{sec:intro}
Character animation has been a central challenge in graphics and vision research, driving innovation across industries from video games and visual effects to emerging virtual and mixed reality experiences \cite{wang2025}. Despite significant progress, achieving high-quality animation—especially for clothing—remains difficult, largely due to the limited compute budgets available on most devices.

Traditional physics-based simulation methods, such as finite element and position-based techniques, are capable of producing realistic cloth motion. However, they suffer from critical drawbacks: high computational cost, lack of robustness, and the need for expert-driven parameter tuning. These factors make real-time performance unattainable on anything but high-end hardware, and full automation for data generation impractical.

Recent learning-based approaches have sought to bridge the gap between realism and efficiency, but they often still require substantial compute and memory resources, or struggle to generalize across diverse body shapes, poses, and garment types. Fast skinning methods, like linear blend skinning (LBS), offer real-time performance but at the expense of physical plausibility, frequently introducing artifacts that break immersion. Hybrid approaches that combine LBS with physics-based simulation in reduced subspaces improve efficiency, yet still fall short of the performance needed for compute-constrained environments.

To address these challenges, we propose a self-supervised learning framework that leverages physics-based losses computed via mesh discretization. This enables us to define stretch and bending losses directly from mesh topology and rest shape, enabling pose generalization without the need for any simulation data. Crucially, while meshes are used for energy evaluation during training, our method is agnostic to garment representation at inference—supporting alternative formats such as Gaussian splats \cite{kerbl2023gs} or point clouds, provided skinning weights are transferable.

Acknowledging the popularity of linear blend skinning for its computational efficiency, despite its notable limitations, we identified a valuable opportunity to improve quality while maintaining compatibility with low compute environments. Furthermore, since most everyday clothing exhibits minimal dynamic behavior during typical avatar movements, effective and efficient quasi-static approximations present a suitable solution. Our neural approach presents a novel modulation based approach to allow for highly efficient inference on CPU devices, yielding substantial qualitative improvements over standard skinning, particularly for loose-fitting garments, enabling visually plausible results and real-time performance on resource-limited devices.

In summary, our method achieves an order-of-magnitude improvement in runtime performance and can even operate on a single-threaded CPU at microsecond inference times, enabling true on-device execution. At the same time, it produces physically plausible garment drapes that generalize across diverse body shapes, poses, and garment topologies, as demonstrated in Figure~\ref{fig:1_teaser}. Our approach advances the state-of-the-art in real-time avatar animation for compute-constrained environments. Our contributions can be summarized as follows:
\begin{itemize}
    \item A lightweight runtime model capable of running on-device with limited compute budget. We demonstrate microsecond performance, orders of magnitude faster compared to prior work, using only widely available single-threaded CPU compute.
    \item A novel hyper-architecture, which exploits a modulation strategy to enable real-time cloth deformation on arbitrary body shapes and poses. Our proposed hyper-architecture is trained with physics-based self-supervision without the need for simulated training data. %\todo{this is arguably not a novel contribution}
\end{itemize}

\section{Related Works}
\label{sec:related_works}

Generating realistic garment deformations under tight compute constraints remains a challenge. While physics-based simulation can produce accurate results for any garment geometry, it is slow and requires extensive manual setup. As a result, there is ongoing research into developing deformation models that are both efficient and generalizable, but achieving this balance continues to be an open problem in the field. % We review cloth dynamics modeling and recent advances in efficient neural models.

\subsection{Physics-based Simulation}

Physics-based methods to simulate highly-detailed clothing is still an active research topic decades after the foundational works~\cite{baraff1998, stuyck2022cloth}. Although capable of generating cinematic quality realism, physics-based methods typically require significant computational resources. To achieve real-time performance for interactive applications, recent research has focused on massively parallel algorithms to leverage GPU hardware~\cite{macklin2016xpbd, chen2024vertex}. Alternatively, simulation has been accelerated by modeling a subspace from which to solve the equations of motion ~\cite{hahn2014subspace, de2010stable, modi2024simplicits, Fulton2018, Chang2023} or predicting latent-space solutions with a neural simulator ~\cite{li2025}.

% Physics-based simulation has long been employed to generate high-quality results in garment modeling~\cite{baraff1998, stuyck2022cloth}. While these methods produce visually realistic outcomes, they typically demand substantial manual setup work and require significant computational resources. To address this, recent research has focused on highly parallel algorithms that leverage GPU hardware for real-time computation~\cite{macklin2016xpbd, chen2024vertex, bouaziz2014projective}.  Despite significant progress, it remains challenging to obtain high quality deformations on low compute devices.

% Segue into "reduced order" in the sense of being specific for garments being worn by avatars.

\newcommand{\no}[0]{{\color{red} \ding{55}}}
\newcommand{\yes}[0]{{\color{green} \checkmark}}
\newcommand{\maybe}[0]{{\color{orange} \nicefrac{1}{2}}}

\begin{table*}[t]
    \centering
    \caption{\textbf{Feature Comparison.} We emphasize that our method possesses several desirable properties and it is significantly faster by orders of magnitude when compared to existing approaches.
    % Any method which only learns one garment or one body shouldn't be called 'independent', they would still need the same topology during inference.
    }
    \resizebox{2.0\columnwidth}{!}{
    \begin{tabular}{rccccccc}
        \toprule
        & SNUG \cite{stantesteban2022snug} & NCS \cite{bertiche2022neural} & HOOD \cite{grigorev2023hood} & VirtualBones \cite{pan2022predicting} & DrapeNet \cite{deluigi2023drapenet} & GAPS \cite{chen2024gaps} & Ours \\
         \midrule
        Self-supervised & \yes & \yes  & \yes & \no & \yes & \yes & \yes \\
        Garment topology independent & \no & \no & \maybe & \no & \yes & \no & \yes \\
        Body topology independent & \no & \no & \maybe & \no & \no & \no & \yes \\
        $>$ 9000 fps on CPU & \no & \no & \no & \no & \no & \no & \yes \\
     \bottomrule
    \end{tabular}}
    \label{tab:feature_parity}
\end{table*}

\subsection{Skinning-based Garment Deformation} 

Alternative to expensive simulation methods, skinning-based methods to deform garments are a common feature in compute-limited environments like video games~\cite{skinningcourse2014}. However, standard linear blend skinning (LBS) can create unappealing garment deformations, as the garment geometry may not map well to the underlying body structure, a problem particularly pronounced with loose-fitting garments. To improve deformation quality over traditional skinning weight methods \cite{Jacobson2011Bounded, larionov2025skincells} optimize skinning weights with physics-based losses and collision penalties over the skinning-based clothing.

% To address modeling garment deformations in a truly compute limited way, researchers and practitioners alike have resorted to simple skinning-based techniques. Standard linear blend skinning  (LBS) methods, which attach clothing to a humanoid skeleton rig, often yield suboptimal results because clothing geometry may not map well to the underlying body structure—this issue is particularly pronounced with loose-fitting garments. To ameliorate this, skinning weights can be optimized to maintain cloth-like properties under deformation~\cite{larionov2025skincells}. 

To further address improved deformations, skinning decomposition~\cite{le2012smooth} extracts the linear blend skinning from a set of examples, effectively modeling the deformation by a low number of rigid bones. Alternatively, learning based methods employ bone-based representation to predict the deformation of loose-fitting garment meshes at interactive rates~\cite{pan2022predicting, halimi2022pattern, li2024ctsn}. More broadly, the concept of virtual bones has been widely adopted to parameterize motion across both graphics and vision. Early deformation and rigging methods~\cite{Sumner2007Embedded, Jacobson2011Bounded, Baran2007Pinocchio} demonstrated that sparse local nodes effectively capture non-rigid deformations—a principle later extended to real-time surface tracking in DynamicFusion~\cite{Newcombe2015DynamicFusion} and LiveCap~\cite{Habermann2019LiveCap}. Recent Gaussian-based representations, including Dynamic 3D Gaussians~\cite{Luiten2023DGS}, RigGS~\cite{Liu2024RigGS}, and Dual Gaussian~\cite{Jiang2024DualGS}, further parameterize complex motions through spatially distributed virtual bones. At the generative level, DiMO~\cite{Mou2025DIMO} distills node-based motion from diffusion models. Collectively, these works demonstrate that virtual bones form a powerful abstraction for parameterizing motion in a variety of representations, including garment deformation.

%\begin{itemize}
%    \item SMPL-based garment reposing/draping (SMPLicit etc.) : directly copying skinning weights, or even %smart transfer of skinning weights from body to garments leads to skinning artifacts. Specifically, they are %not physically plausible. (Can we also mention skin cells here)?
%    \item DEM-bones : defining bones for garments and learning skinning weight distribution : fit to one given %garment and animation. 
%    \item Virtual bones (learn pose-conditioned and material conditioned deformation over DEM-bones) limited to %a single body shape and single garment, expensive during inference, supervised training : requires expensive %simulation data.
%    \item (ClothFiT should also come here somewhere)
%\end{itemize}

\subsection{Neural Garment Simulation}

In recent years, two primary approaches have emerged for learning garment simulation: supervised models relying directly on data, and self-supervised models relying on physics-based losses instead of data. Supervised methods like ~\cite{holden2019subspace} combine subspace simulation techniques with machine learning to enable efficient subspace-only physics simulation. TailorNet~\cite{patel2020tailornet} predicts deformations in frequency space and models clothing dynamics conditioned on pose, shape, and style. Others \cite{li2024neural} leverage a manifold-aware transformer framework to predict deformations. However, these methods require simulation data that is often difficult and expensive to obtain. In contrast, self-supervised approaches have shown effectiveness in modeling physical systems without the need for precomputed training data~\cite{stantesteban2022snug, bertiche2022neural, stuyck2025quaffure, lin2025neuralocks}. SNUG~\cite{stantesteban2022snug} introduced a self-supervised training method for modeling dynamic clothing deformations in a data-free approach. NCS~\cite{bertiche2022neural} extended this work by devising an architecture able to automatically disentangle static and dynamic cloth subspaces, improving quality of the results. DrapeNet~\cite{deluigi2023drapenet} presented a strategy enabling it to generalize across garments for modeling quasi-static deformations. HOOD~\cite{grigorev2023hood} presented a graph neural network based approach to model cloth dynamics that generalizes across body and garments shapes for varying poses. Follow-up work improved performance~\cite{li2024efficient}. Graph neural network have also been successfully employed to enable cloth upsampling~\cite{halimi2023physgraph, zhang2024neural}.

Despite notable progress, existing methods remain computationally prohibitive for low-resource devices, as each forward pass relies on GPU, transformer, or graph-based deformation models. Furthermore, these approaches are limited by the requirement of accurate initial draping of the garment in a canonical shape.

%\begin{itemize}
    %\item Supervised Methods: TaylorNet, , ManifoldAwareTransformers : generalize to different body shapes and garments, but require expensive simulation data  to train. 
    %\item Self-supervised Methods: DrapeNet~\cite{deluigi2023drapenet} \& DIG (only quasi-static, SMPL-based so very restrictive). SNUG~\cite{stantesteban2022snug} (handle dynamics and generalizes to different parametric shapes and poses, restricted to SMPL), GAPS (better skinning weight initialization for each garment vertex (rbf over SMPL skinning weights)) -- better results than SNUG but learns one-garment per training. HOOD - self-supervised, generalizes across, shapes, garments and poses, handle dynamics and friction as well -- biased to SMPL triangulation.
    %\item Efficient Deformation Learning of Varied Garments with a Structure-Preserving Multilevel Framework \cite{li2024efficient}
    %\item All of the simulation methods are expensive to infer due to either GPU, transformer or graphnets-based deformers being used in every forward pass. Also, all of them required good initial draping of the garment in canonical shape. None of them generalize to arbitrary bodies (we don't also yet, but we'll see about that and omit accordingly).
%\end{itemize}

\begin{figure*}[t]
    \centering
    \includegraphics[width=0.92\linewidth]{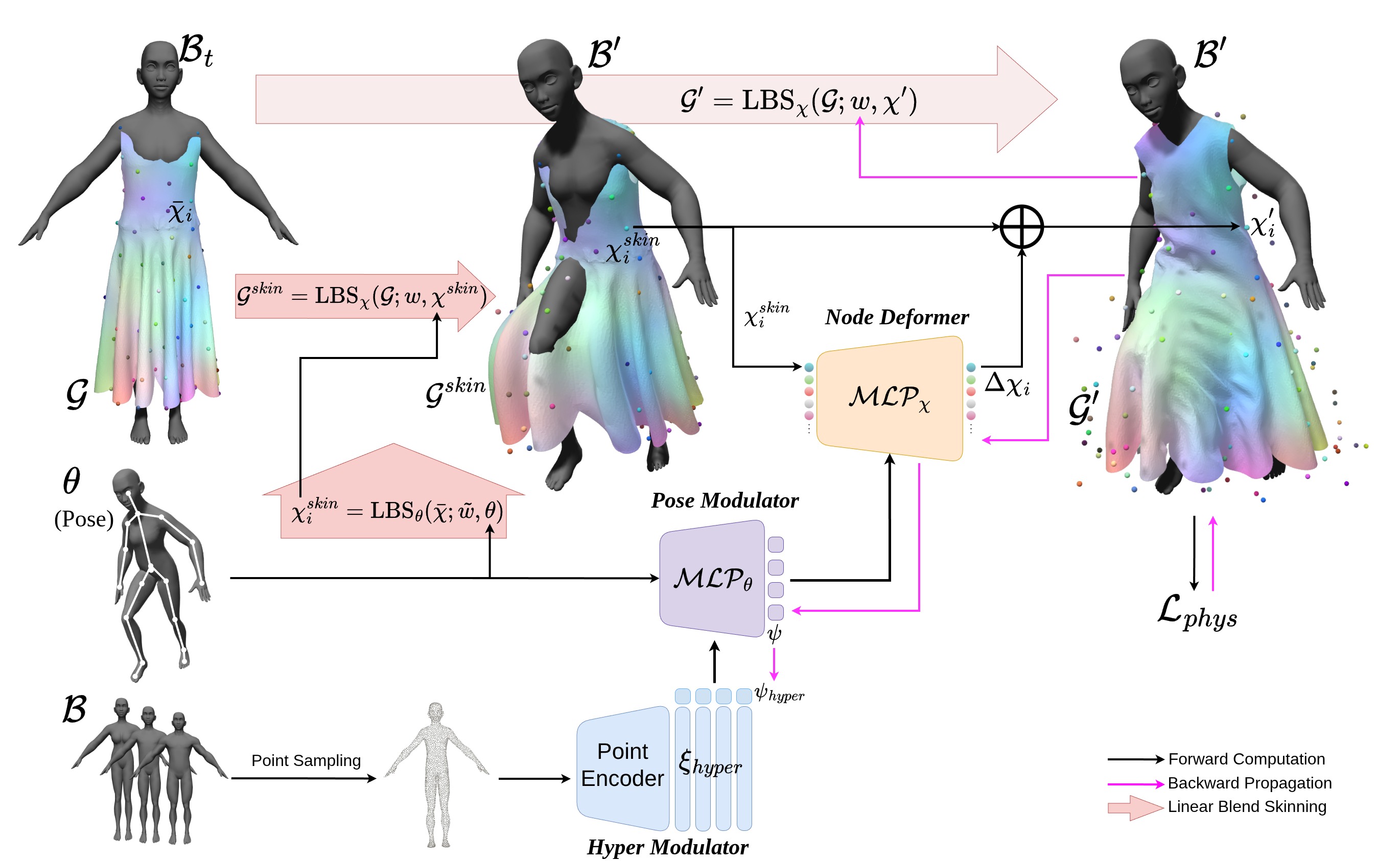}
\caption{\textbf{Method overview}. Given a garment $\G$ and pose $\theta$, our method first applies a sequence of two LBS computations to predict the location of the garment for the particular pose indicated by the wide arrows. 
Given the body geometry $\B$, we sample its surface with points and compute a modulation signal using a transformer network $\Hyper$ for $\MLPt$. The Pose Modulator and Node Deformer networks work in conjunction to predict the best corrective for the LBS to predict the most optimal drape as measured by the physically based loss $\mathcal{L}_{\mathit{phys}}$.}
\label{fig:pipeline}
\vspace{-0.3cm}
\end{figure*}

\subsection{Universal Encoding for Clothed Humans}
Early approaches to clothed human modeling relied on subject-specific optimization of implicit or parametric representations, which limited generalization to unseen identities. 
Methods such as SMPLicit~\cite{corona2021smplicit} and ICON~\cite{xiu2022icon} introduced canonical implicit functions that jointly encode the body shape and clothing geometry within a shared space, enabling pose transfer and coarse garment preservation. 
Subsequent work such as Neural-GIF~\cite{tiwari2021neural} or VS~\cite{liu2024vs} explored learned canonical mappings and vertex-aligned features  to improve geometric fidelity and disentangle pose from identity. Recent advances shift toward universal, data-driven encoders that jointly model diverse body shapes and garment topologies without test-time optimization. 
FRESA~\cite{wang2025fresa} and ReLoo~\cite{guo2024reloo} focus on training large-scale networks over thousands of subjects to learn clothing-aware deformation fields and disentangled skinning weights that generalize across identities and motion. 
Another line of work~\cite{kim2024gala, vuran2025remu} decouples the clothed human into layered assets (\ie, body, clothing, and accessories) allowing compositional generalization and garment re-targeting. 
Other feed-forward pipelines~\cite{ye2025freecloth,jin2025monocloth}, leverage volumetric or Gaussian representations to capture complex fabric dynamics within a unified latent space. Collectively, these methods demonstrate a shift from optimization-based reconstruction toward learned universal encoders that explicitly factorize geometry, pose, and appearance—laying the foundation for scalable, animatable digital humans that handle the full variability of real-world clothing.
Finally, rig-prediction approaches~\cite{zhang2025unirig,xu2022morig,RigNet} learn to predict joint skeletons and skinning weights for new characters but do not explicitly model pose-dependent surface deformation or garment dynamics.

% Material prediction, then layered prediction.
% For universal conditioning, FRESA, GALA, 
% on topics relating to how learning techniques are used to effectively encode garments and body shapes.
% \begin{itemize}
%     \item ReLoo
%     \item Smplicit: Topology-aware generative model for clothed people
% \end{itemize}

% \el{available references:}
% \begin{itemize}
%     \item SIREN: \cite{sitzmann2020siren}
%     \item Modulated-SIREN: \cite{mehta2021}
% \end{itemize}

% GALA: Generating Animatable Layered Assets
% Puppeteer: RigandAnimate Your 3D Models
%  Smplicit: Topology-aware generative model for clothed people

% \newcommand{\B}[0]{\mathcal{B}}
% \newcommand{\G}[0]{\mathcal{G}}
% \newcommand{\Gs}[0]{\mathcal{G}^{\mathit{skin}}}
% \newcommand{\R}[0]{\mathbb{R}}
% \newcommand{\MLPx}[0]{\mathcal{MLP}_{\chi}}
% \newcommand{\MLPt}[0]{\mathcal{MLP}_{\theta}}
% \newcommand{\hyper}[0]{\mathit{hyper}}
% \newcommand{\Hyper}[0]{\xi_\hyper}
% \newcommand{\psih}[0]{\psi_\hyper}

\section{Method}
\label{sec:method}

Given a garment mesh draped (or roughly aligned) on a canonical body shape and pose, our draping framework produces physically plausible deformations, generalizing across varying body shapes and poses. An overview of our method is illustrated in \figref{fig:pipeline}. What distinguishes our approach from previous work is its optimization for rapid inference on low-compute devices. By decoupling body pose and shape through modulation of the node deformation network, we are able to maintain compact neural networks, resulting in exceptionally fast inference times. Furthermore, our use of self-supervised learning eliminates the need for costly simulation data during training. Our skinning-based framework is lightweight and versatile: once trained, the network can drape the given garment with different resolutions and topological variations across a wide range of body shapes and poses, making it highly suitable for real-world applications on edge devices.

%\textbf{TODO we need a paragraph about what sets our method apart. Separate pose prediction from body shape... The model design MLPx allows for efficient inference todo explain how why}

% To achieve real-time performance, we first approximate the coarse deformation of the garment using the linear blend skinning (LBS) method, which requires minimal compute cost and memory overhead. Then to 
% Our method achieves real-time performance for high-quality garment deformation by first approximating the coarse deformation with the extremely fast method of linear blend skinning (LBS), then applying a pose-based neural corrective using a small, efficient neural network. To achieve generalization over bodies we employ a point transformer architecture to modulate the aforementioned pose-conditioned network.

\subsection{Inputs}

The input is composed of three components. (1) Garment geometry
$\G$ is a triangle mesh in canonical space. 
% This is sufficient to model garment physics losses in our exposition, however, in practice, our method can also make use of a separate 2D panel mesh with the same number of triangles to compute more physically accurate forces in the loss, as one would typically employ in a standard simulator \cite{stuyck2022cloth}.
(2) Body geometry, $\B$, is a similarly defined mesh. (3) Pose $\theta \in \R^p$ (in our case, $p=120$) is a low-dimensional parametrization of the joint state (affine transforms of the joints) of a common hierarchical humanoid skeleton rig with 128 bones that binds $\B$ with predetermined skinning weights. Our method applies to other rigs as well. 

%In practice, we use a constant manually set linear transform to map $\theta$ to the set of all joint angles and bone lengths, but in the remainder of this section we use joint state interchangeably with $\theta$.

\subsection{Node-based Skinning for Garments}

Garment deformation is driven by a set of $m$ affine handles or nodes, defined by $\chi_i \in \R^{4\times4}$.
% In our experiments, we found $m=128$ nodes strikes a balance between deformation quality and computation time. 
The $n$ vertices of the garment are bound to nodes via the skinning weight matrix $w\in\R^{m\times n}$, such that each vertex $x \in \G$, represented in homogeneous coordinates, is transformed as
% \todo{TO DO explain $\bar{\chi}_i^{-1}$}
\begin{align*}
    \LBS_\chi(x; w, \chi) & = \sum_{i=1}^m w_{i} \chi_i \bar{\chi}_i^{-1}x \quad \text{where} \quad \sum_{i=1}^m w_{i} = 1.
\end{align*}
% \todo{summarize \cite{skinningcourse2014} for skinning and part of \cite{pan2022predicting} specifically on representing garment deformation with virtual bones}
Skinning weights are initialized as 
\begin{align*}
    w_{i,j} & = \exp(-d_{i,j}/\sqrt{n/m})
\end{align*}
where the distance $d_{i,j}$ between node $i$ and vertex $j$ is computed using the heat method \cite{crane2017heat}. The set of nodes is sampled using a farthest point strategy on each garment mesh in canonical pose giving the last column of $\bar \chi$. The linear part of $\bar \chi$ is set to the identity.

The resulting sampled nodes $\chi$ form a reduced degrees-of-freedom representation of the garment geometry, enabling a simulator or learned model to efficiently deform the garment through these nodes. Additionally, training a model to deform the garment nodes $\chi$ from canonical space to posed space directly would be unnecessarily challenging. To simplify, we first initialize $\chi$ from joint states $\theta$ using a separate LBS computation. Formally, $\chi^{\mathit{skin}} = \LBS_\theta(\chi; \tilde{w}, \theta)$ where $\tilde{w}_{i,k}$  is the inverse squared distance from each garment node $i$ to skeleton bone $k$ if $k$ is the closest, and 0 otherwise. This operation results in an initial transformation of the garment nodes, and thereby of garment vertices from canonical space to pose space.

% Node-based skinning allows our method to support arbitrary mesh resolutions and cropped versions of garments that the architecture is trained on since we simply need to recompute the mapping from the predicted nodes to the specific topology or representation at hand. The rest of the section will cover the details of adding neural correctives $\Delta\chi$ to $\chi$.

\subsection{Network Architecture}

User customization of their avatars is an important aspect of creativity and immersion in interactive applications. As such, generalizing garment deformation across body shapes is essential to prevent expensive retraining for every user. While it is possible to use a monolithic network to directly map body shape $\B$ and pose $\theta$ to node transformation deltas $\Delta\chi$, this approach is unnecessarily expensive at runtime—especially since only $\theta$ changes during inference. In such a setup, the network would need additional capacity for interpreting the body shape signal for every evaluation, leading to inefficient memory and compute costs.

Instead, we adopt a modulation strategy to efficiently pass the body and pose signal. This design decision allows us to keep our networks small such that they map well to widely available CPU compute. Modulation has been shown to yield improved results in domains such as image encoding~\cite{mehta2021}, outperforming multi-layer perceptrons with input-concatenated conditioning. Our method is partitioned into three components: (1) the Hyper Modulator interprets the body shape $\B$ and generates a modulation signal $\psih$, (2) the Pose Network ingests $\psih$ to modulate its activations, while otherwise depending only on the current pose $\theta$, and (3) the Node Deformer is subsequently modulated by the Pose Modulator output, enabling it to predict node transformation deltas $\Delta\chi$ for a body shape and pose efficiently and accurately.

\noindent\textbf{Hyper Modulator}: 
Given a body mesh $\B$ in canonical pose, we sample 4096 points and their corresponding normals, apply positional encoding, and feed these point tokens to a standard transformer encoder layer with 4 attention heads comprising $\Hyper$. This architecture and point-sample tokenization makes our method independent of the mesh topology. The point tokens are then further pooled via Self-Attention to obtain a global body shape conditioning signal, which is then fed into residual-MLP blocks (with skip connections). The final output is a set of modulation signals that scale and shift the activation layers of the $\MLPt$.

\noindent\textbf{Pose Modulator}: 
Given a pose $\theta$, the pose modulator $\MLPt$ computes another modulation signal using a standard MLP (4 hidden layers, each with 512 neurons), which is then fed into the Node Deformer.

\noindent\textbf{Node Deformer}:
The Node Deformer, $\MLPx$ is also an MLP with 4 hidden layers and 128 hidden features, being modulated by $\MLPt$.  $\MLPx$ is responsible for computing the per-node corrective transformations for the garment nodes. It takes as input a concatenation of all posed node transformations $\left(\chi_i^{skin} \hspace{4pt} \forall i\right)$ and produces the appropriate correctives.

% In addition to the signal received from the pose modulator, this input enables the network to process additional signal related to the pose. The Node Deformer uses 4 hidden layers with 128 hidden features each.

\subsection{Training}\label{sec:training}

During training we optimize the skinning weights $w$ and the network weights for $\MLPx$, $\MLPt$ and $\Hyper$ by evaluating physics-based losses in a self-supervised setting. We sample poses $\theta$ uniformly within a bounded region centered around the A-pose. Using a set of manually defined joint limits $(\theta_{\text{min}},\theta_{\text{max}})$ roughly corresponding to realistic anatomical limits. We gradually sample with increasingly larger bounds starting from $(0.1\theta_{\text{min}}, 0.1\theta_{\text{max}})$ and increasing by $(0.1\theta_{\text{min}}, 0.1\theta_{\text{max}})$ for every subsequent 100 epochs. This strategy allows us to avoid large contact penalties during training early on due to poor initial drape results, which stabilizes the learning progress.

\noindent\textbf{Loss Functions}:
We use physics-based losses during training to ensure the final, corrected garments are physically-plausible. Cloth elasticity is modeled with St. Venant-Kirchhoff in-plane energy $E_{\text{StVK}}$~\cite{barbivc2005stvk} and dihedral bending energy $E_{\text{dihedral}}$~\cite{grinspun2003shells}. To avoid interpenetration, we penalize cloth vertices within 0.3 cm of the surface of the body, using a collision penalty $E_{\text{collision}}$ proportional to the squared interpenetration distance. We use a simple gravity energy loss computed as $E_{\text{grav}} = \sum_k g m_k y_k$ where  $m_k$ is the mass computed from the density of the fabric, $y_k$ the height of vertex $k$ and $g= 9.81$ m/s${}^2$. 
The total loss is the sum of physical energy components: 
\begin{align}
\mathcal{L}_{\text{phys}} = \lambda_1 E_{\text{StVK}} + \lambda_2E_{\text{dihedral}} + \lambda_3E_{\text{collision}} + \lambda_4E_{\text{grav}} \label{eq:loss}
\end{align}
where $\lambda_1$,$\lambda_2$,$\lambda_3$, and $\lambda_4$ are the scaling terms precomputed based on standard material properties.
%This avoids garments incorrectly passing through thin body features such as arms.

% \begin{table}[t]
%     \centering
%     \caption{\textbf{Material parameters} used to compute \myeqref{eq:loss}.}
%     \resizebox{0.5\columnwidth}{!}{
%     \todo{
%     \begin{tabular}{l c}
%         \toprule
%          Material parameter & Value \\
%         \midrule
%         Young's modulus & 1 \\
%         Poisson's ratio & 2 \\
%         Bending stiffness & 3 \\
%         Collision penalty coeff. & 4 \\
%         \bottomrule
%     \end{tabular}
%     }}
%     \label{tab:matparam}
% \end{table}

\subsection{Inference}

Given a new body geometry, we begin by passing a sampled set of points to the hyper modulator $\Hyper$, which generates modulation signals for the pose modulator $\MLPt$. This step is performed only once per body shape. During runtime, each incoming pose vector is efficiently mapped to a deformed garment by applying linear blend skinning (LBS) with pose-modulated garment handle correctives. This approach enables fast and accurate garment deformation for varying poses relying on CPU compute only.
\begin{table}[t]
    \centering
    \caption{\textbf{Inference time} (in microseconds) using our method. Results are shown for both PyTorch GPU and single-threaded CPU execution for a custom C++ implementation of the forward pass, demonstrating that the method enables true on-device performance. The reported timings correspond to the evaluation of the two MLPs required per frame. Note that these numbers do not include the skinning computation nor the body encoder, which only needs to be computed once per body type and is not part of the per-frame inference cost. CPU measurements were taken on an AMD Ryzen Threadripper PRO 7995WX.}
    \resizebox{1.0\columnwidth}{!}{
    \begin{tabular}{l c c c}
        \toprule
         & $\MLPx$ ($\upmu$s) & $\MLPt$ ($\upmu$s) & FPS \\
        \midrule
        CPU (C++, single-threaded) & 12.5 & 46.7 & 16,891 \\
        % GPU (PyTorch) & \todo{x} & \todo{x} & \todo{x} \\
        \bottomrule
    \end{tabular}}
    \label{tab:performance}
\end{table}

\section{Experiments}
\label{sec:exp}

Here we summarize implementation details and the datasets used, and present ablation studies and comparisons with related methods.

\subsection{Implementation} \label{sec:impl}
We implemented all internal components of the model in PyTorch~\cite{paszke2019pytorch}, and physics-based losses comprising $\mathcal{L}_{\text{phys}}$ in NVIDIA Warp~\cite{warp2022}. Training takes approximately $32$ hours on a single NVIDIA A100 GPU. We developed a custom C++ implementation of the MLP evaluation (both $\MLPx$ and $\MLPt$) to enable efficient CPU usage at runtime using the weights optimized during training. Inference timings are reported in Table~\ref{tab:performance}.

\subsection{Dataset} \label{sec:dataset}
For training shape hypermodulator, we use an in-house SMPL-like parametric body model, with richer shape diversity and additional surface impressions (e.g. muscles). Please note that, unlike existing methods, for model-agnosticism, we only sample the surface shape information for body shape encoding, and do not rely on PCA blend-shapes, etc. We sample poses $\theta$ from AMASS dataset and use the same split as HOOD\cite{grigorev2023hood} (as described in~\secref{sec:training}). No simulated garment data is needed due to our self-supervised setting. We sample garments for both evaluation and training from GarmentCodeData\cite{korosteleva2024gcd}. \\
To further show the ability of our framework to adopt to diverse body shape, we train our framework separately on 3DBiCar \cite{luo2023rabit} dataset containing fantastical characters and arbitrarily proportioned body shapes. We show results on 3DBiCar in canonical pose in the teaser figure. Since, the authors haven't released the motion data with 3DBiCar, we don't train across motion sequences. Please refer to suppl. for more qualitative results.

\subsection{Ablation Studies} \label{sec:ablations}
\paragraph{Error Metrics} We follow GAPS \cite{chen2024gaps} for reporting strain errors $\epsilon_e$ and $\epsilon_a$ in the following sections measuring the mean difference (in \%) between edge lengths and areas between the template and draped garment. We also employ $\epsilon_c$ as the \% of draped garment vertices colliding with the body.

\paragraph{Garment Node Resolution} We evaluate the quality-cost tradeoff of increasing the number of garment nodes during training. The metrics presented in Table~\ref{tab:ablation} indicate a diminishing return on quality after 128 nodes but a significant increase in cost. Therefore, we chose to use 128 garment deformation nodes in our method.

% \paragraph{Number of Nodes} We evaluate our method trained with different numbers of nodes (Table~\ref{tab:ablation}) to indicate the quality improvement as the number of nodes increases. The metrics don't show a significant improvement after the number of nodes is increased past 128 indicating that our choice of discretization is a good balance between performance and quality.  
\begin{table}[t]
    \centering
    \caption{\textbf{Ablation Studies.} In the first row, we demonstrate the effect of removing skin weight optimization from the training process, indicating must higher strains and collision rates across the board. The second section shows the effect of training with fewer nodes, showing lower metrics with diminishing differences as the number of bones is increased. This supports our choice for LBS-based discretization for garments indicating that simulating the full mesh is often unnecessary.}
    \resizebox{0.9\columnwidth}{!}{
    \begin{tabular}{c c c c c c }
        \toprule
        \# of Nodes & Skin Weight Opt. & $\epsilon_e$ $\downarrow$  & $\epsilon_a$ $\downarrow$ & $\epsilon_c$  $\downarrow$ & $LBS$ overhead ($\upmu$s)  $\downarrow$\\
        \midrule
        32 & \no & 7.010 & 8.091 & 0.316 &  +2.20\\
        64 & \no & 5.817 & 6.103 & 0.207 &  +4.46\\
        128 & \no & 4.266 & 4.093 & 0.129 &  +9.03\\
        \midrule
        64 & \yes & 4.512 & 4.718 & 0.126 & +4.46 \\
        \cellcolor{blue!25}{128} & \cellcolor{blue!25}{\yes} &  \cellcolor{blue!25}{3.074} &  \cellcolor{blue!25}{3.193} &  \cellcolor{blue!25}{0.119} & \cellcolor{blue!25}{+9.03}\\
        256 & \yes & 3.068 & 3.188 & 0.117 & +18.11  \\
        \bottomrule
    \end{tabular}}
    \label{tab:ablation}
\end{table}

\paragraph{Skinning Weight Optimization} Our scheme exploits the ability of skinning weights to improve the quality of deformation through skinning weight optimization \cite{larionov2025skincells}. We show how skinning weight optimization can improve the results quantitatively in the first row of Table~\ref{tab:ablation}.

\begin{figure}[h!]
    \centering
    \includegraphics[width=\linewidth]{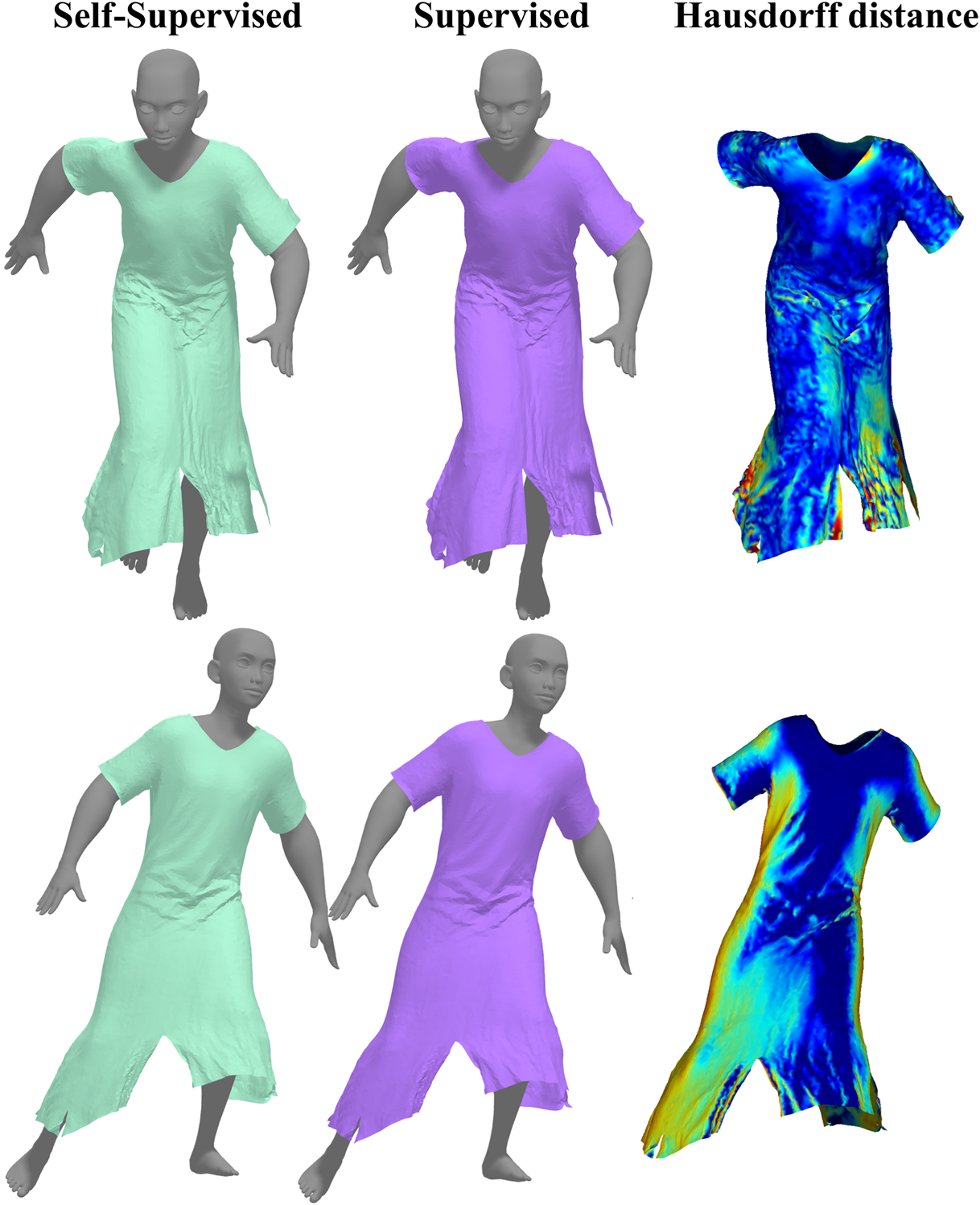}
    \caption{Comparison of Self-Supervised Training with Supervised Training.
    Our proposed self-supervised training scheme does not require additional simulation software and yet can achieve the same result as when using supervision. On the right, we map the Hausdorff distance between the two results indicating that they are indeed slightly different.
    % \todo{It's a little suspicious that they are this close, maybe describe how the presimulated results were obtained.}
    % \todo{Do we even need the error plot, if we do we should add legend}
    % \todo{In MeshLab, you can create a difference plot (a heatmap of distances) between two aligned meshes using the "Distance from reference mesh" or "Hausdorff Distance" filters. This process involves calculating per-vertex distances and then visualizing them as a color gradient. [Todo: Astitva]}
    }
    \label{fig:virtualbones}
\end{figure}

\paragraph{Supervision} Taking a single garment and body, we run a series of quasi-static simulations for a sampling of $1$k poses to build a sample simulation dataset. For each pose we take 1000 steps of the Adam optimizer over garment vertices, directly minimizing the physics loss \myeqref{eq:loss} to mimic a typical simulation system. We use this set to train our architecture (using an L2 loss in place of \myeqref{eq:loss}) between the deformed and simulated garment vertices to demonstrate supervised training. 
In \figref{fig:virtualbones} we demonstrate that a similar result can be generated, purely using our self-supervised training strategy proposed in \secref{sec:training}.

\subsection{Comparisons}

\begin{table}[t]
    \centering
    \caption{\textbf{Drape Quality Metrics.} A comparison of GAPS, HOOD, and our method shows that our approach is most effective at eliminating interpenetration, which leads to the most noticeable artifacts. Removing these intersections is advantageous, even if it results in a slight increase in strain metrics. For strain metrics $\epsilon_e$ and $\epsilon_a$, our method performs comparably to HOOD, while being orders of magnitude faster to evaluate.}
    \resizebox{0.9\columnwidth}{!}{
    \begin{tabular}{lrccc}
        \toprule
         \textbf{Method} & $\epsilon_e$ $\downarrow$  & $\epsilon_a$ $\downarrow$ & $\epsilon_c$  $\downarrow$\\
         \midrule
         GAPS~\cite{chen2024gaps} (1 Step) & 4.231 & 4.655 & 0.153 \\
         HOOD~\cite{grigorev2023hood} (1 Step) & 6.482 & 5.993 & 0.182 \\
         HOOD~\cite{grigorev2023hood} (10 Steps) & \cellcolor{green!25}{2.987} & \cellcolor{green!25}{3.061} & \cellcolor{yellow!25}{0.125} \\
         \midrule
         Ours (1 Step) & \cellcolor{yellow!25}{3.074} & \cellcolor{yellow!25}{3.193} & \cellcolor{green!25}{0.119}  \\
         \bottomrule
    \end{tabular}}
    \label{tab:comparisons}
\end{table}
We compare our method with popular methods providing the capability of draping garments on arbitrarily posed body geometries. First we compare our approach to others in terms of features in Table~\ref{tab:feature_parity}. Our method is the only known to us that enables body and garment topology changes at inference time, while maintaining the convenience of self-supervised training. Most importantly our method achieves a far superior frame rate on CPU, making it easily deployable to compute constrained environments.

We closely investigate the quantitative differences in drape quality in Table~\ref{tab:comparisons}, comparing our method further against HOOD and GAPS. These are a natural choice since they outperform other methods like SNUG. VirtualBones is the most similar to our Supervision ablation in \secref{sec:ablations}, where our method shows comparable quality, yet at a fraction of the performance cost. DrapeNet requires a latent code for an arbitrary garment for a pretrained garment autoencoder, which limits the types of garments it can be compared against. One caveat with GAPS and HOOD is that they model dynamics, which may not produce a valid drape unless rolled out with enough steps. GAPS can produce a drape by being fed the same pose spanning the entire roll-out window, whereas HOOD requires multiple steps since it uses previous cloth state to predict the next similar to a classical simulator. We found that with HOOD 10 roll-out steps is enough to produce a reasonable drape approximation for comparison purposes.

\begin{figure}[h!]
    \centering
    \includegraphics[width=\linewidth]{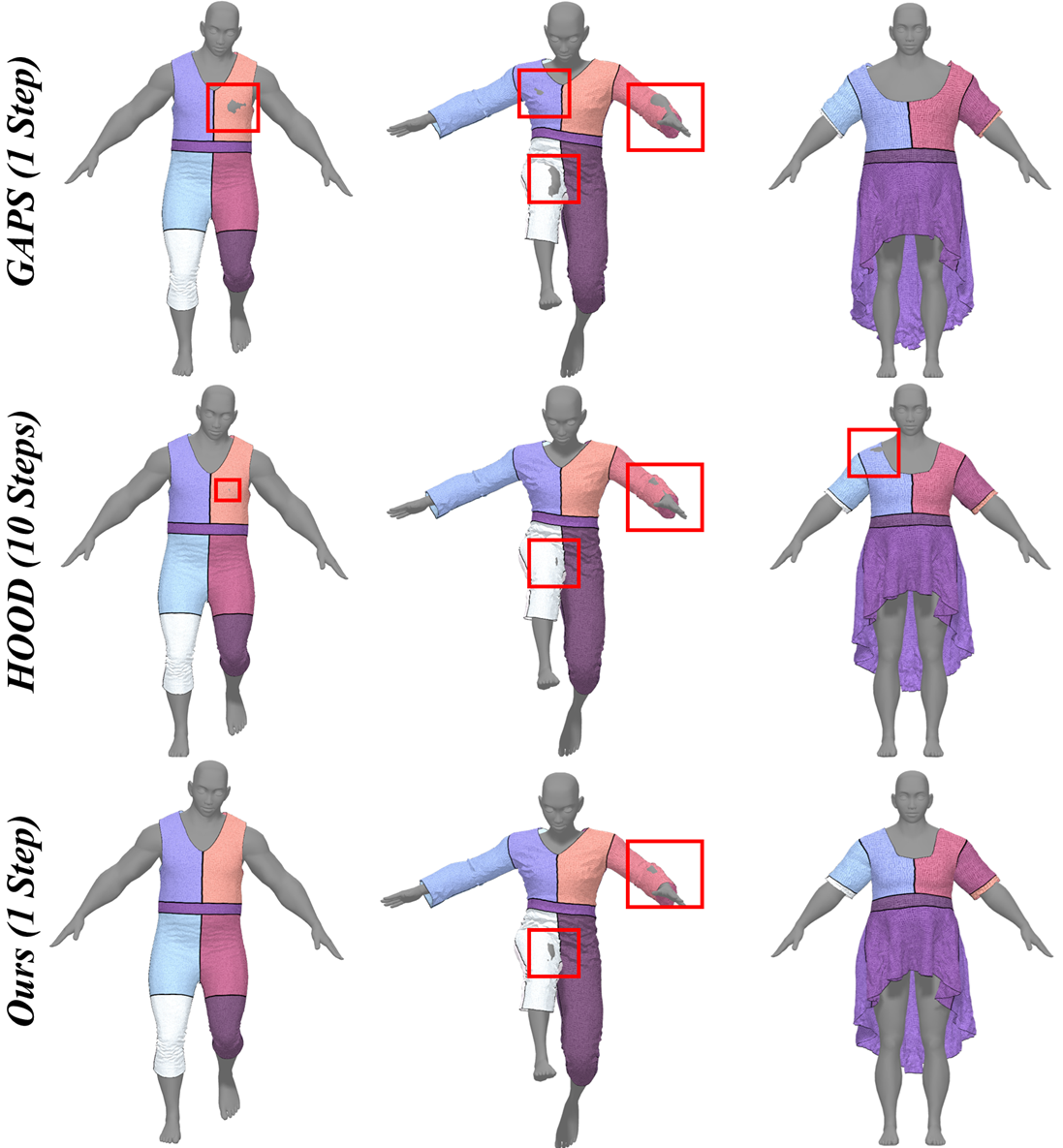}
    \caption{\textbf{Qualitative Drape Comparison}. Our method is qualitatively compared against draping results from GAPS and HOOD on a variety of body shapes and poses. Here we demonstrate that our method produces comparable quality to both GAPS and HOOD, with slightly better overall collision handling, and all at a fraction of the cost.
    % \todo{Add conclusion, how are we better? Better generalization? Better performance? Can we show more obvious body shape variation and more pose variation?}
    }
    \label{fig:hood-gaps}
\end{figure}

\paragraph{Drape Quality}
In \figref{fig:hood-gaps}, we show a qualitative comparison against methods that support arbitrary body topologies like GAPS and HOOD. There we see a comparable quality with slightly better collision handling from our method, but at the fraction of the cost.

We also compare drape quantitatively in Table~\ref{tab:comparisons}, leveraging the same metrics from \secref{sec:ablations}. Here we run our method, GAPS and HOOD on a test set comparable to the AMASS CMU test split to generate an aggregate metric indicating average strain and collision rate. The results indicate that our method outperforms GAPS in all metrics and is comparable with HOOD, but at small fraction of the inference cost.

\begin{figure}[h!]
    \centering
    \includegraphics[width=\linewidth]{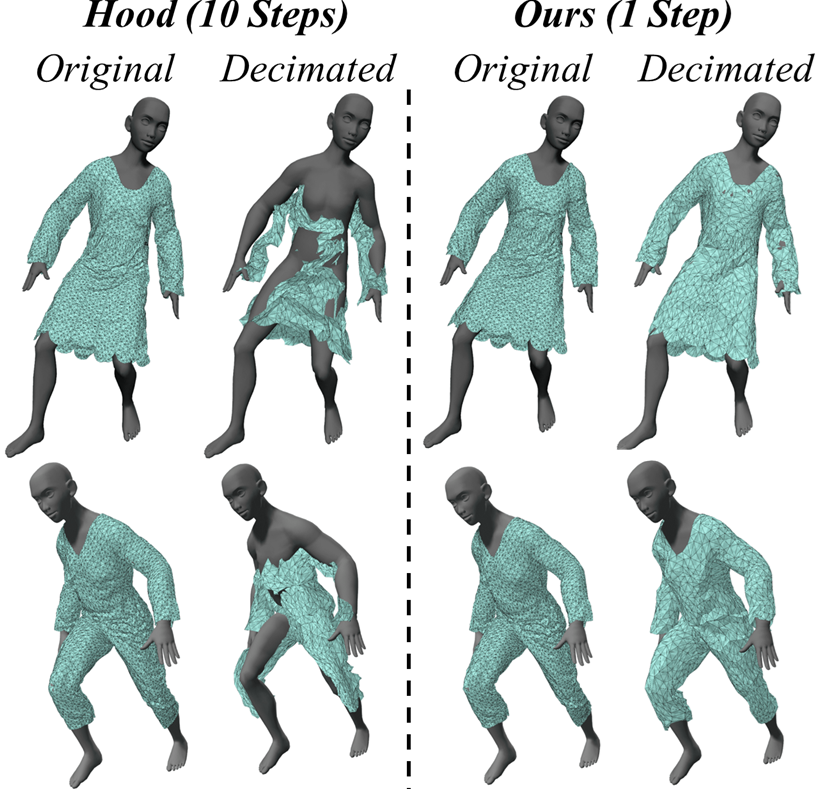}
    \caption{\textbf{Resolution Agnostic Draping}. We demonstrate that our method is able to produce physically plausible drapes for varying resolutions of the garment it was trained on.  In contrast, HOOD fails to produce an adequate drape when applied to mesh resolutions that differ significantly from those it was trained on.}
    \label{fig:resolution}
\end{figure}

\paragraph{Garment Remeshing}
Our method supports dynamic garment topology changes, enabling adaptive remeshing, dynamic level-of-detail or mesh refinement. While HOOD has some generalizability to different meshing, nature of graph neural networks employed within cause it to fail for meshes with sufficiently different edge length. In \figref{fig:resolution}, we demonstrate how decreasing mesh resolution causes no problems for our method, but fails catastrophically for HOOD. 

%Compare competing methods in some quantitative metrics. For example use the same energy model for training data in different methods but use quasi-static data only. If this works we have one solid point of reference for quantitative comparison.

% \newcommand{\no}[0]{{\color{red} \ding{55}}}
% \newcommand{\yes}[0]{{\color{green} \checkmark}}
% \newcommand{\maybe}[0]{{\color{orange} \nicefrac{1}{2}}}

% \begin{table*}[t]
%     \centering
%     \caption{\textbf{Feature Comparison.} We emphasize that our method possesses several desirable properties and it is significantly faster by orders of magnitude when compared to existing approaches.
%     % Any method which only learns one garment or one body shouldn't be called 'independent', they would still need the same topology during inference.
%     }
%     \resizebox{2.0\columnwidth}{!}{
%     \begin{tabular}{rccccccc}
%         \toprule
%         & SNUG \cite{stantesteban2022snug} & NCS \cite{bertiche2022neural} & HOOD \cite{grigorev2023hood} & VirtualBones \cite{pan2022predicting} & DrapeNet \cite{deluigi2023drapenet} & GAPS \cite{chen2024gaps} & Ours \\
%          \midrule
%         Self-supervised & \yes & \yes  & \yes & \no & \yes & \yes & \yes \\
%         Garment topology independent & \no & \no & \maybe & \no & \yes & \no & \yes \\
%         Body topology independent & \no & \no & \maybe & \no & \no & \no & \yes \\
%         $>$ 9000 fps on CPU & \no & \no & \no & \no & \no & \no & \yes \\
%      \bottomrule
%     \end{tabular}}
%     \label{tab:feature_parity}
% \end{table*}

\section{Limitations and Future Work}\label{sec:future_work}

\paragraph{Generalization} We have shown that our proposed architecture can generalize over body shape and pose for a given garment to produce drapes. However, full generalizability is desirable for unlocking a completely automated pipeline where clothing can be immediately deformed. Recent advances in generative techniques have dramatically improved garment creation, allowing users to design new clothing from simple text prompts, scans, or images \cite{chen2025dressanyone, srivastava2025wordrobe, li2024diffavatar, sarafianos2025garment3dgen}. Enabling garment deformation without the need for garment specific models will unlock numerous applications.

\paragraph{Dynamics} Our method produces plausible draping on arbitrary poses. However, many real-world applications like video games or dynamic virtual reality experiences involve quick motions that ought to trigger dynamic garment behavior. Dynamics in auto-regressive neural simulators is an active research area~\cite{li2025, holden2019subspace}. While we believe that prior work~\cite{stantesteban2022snug, bertiche2022neural, lin2025neuralocks} can be leveraged to extend our pose modulator network to handle dynamics, making a significant advance in this direction warrants a dedicated study.

\section{Conclusion}
\label{sec:conclusion}

We introduced a novel and efficient method for generating draped garment configurations across diverse body shapes and poses. Our approach leverages fully self-supervised training, removing the dependency on costly simulation data. The resulting networks are highly flexible, supporting various garment resolutions and representations—even those with modified topologies. We achieve microsecond inference times by proposing a new strategy that modulates bone predictions based on pose and body shape. This allows us to significantly reduce network size, enabling efficient inference times and true on-device performance by leveraging only widely available CPU compute. Through comprehensive experiments, we show that our method consistently outperforms state-of-the-art approaches, enabling garment draping in compute-constrained environments.

\clearpage
{
    \small
    \bibliographystyle{ieeenat_fullname}
    \bibliography{MAIN}
}

\end{document}